# Enhancing Transparency and Traceability in Healthcare AI: The AI Product Passport

| Title | **Enhancing Transparency and Traceability in Healthcare AI: The AI Product Passport** |
|---|---|
| **Authors** (highest academic degree) (ORCID) | 1. A. Anil Sinaci[a] (PhD) (0000-0003-4397-3382)<br>2. Senan Postaci[a] (MSc) (0000-0001-8835-242X)<br>3. Dogukan Cavdaroglu[a] (BSc) (0000-0001-7033-3116)<br>4. Machteld J. Boonstra[b] (PhD) (0000-0001-7550-0489)<br>5. Okan Mercan[a] (BSc) (0009-0005-5178-8592)<br>6. Kerem Yilmaz[a] (BSc) (0009-0005-3725-560X)<br>7. Gokce B. Laleci Erturkmen[a] (PhD) (0000-0002-6201-3849)<br>8. Folkert W. Asselbergs[b,c,d] (Professor) (0000-0002-1692-8669)<br>9. Karim Lekadir[e,f] (Professor) (0000-0002-9456-1612) |
| **Affiliations** | - [a]SRDC Software Research Development and Consultancy Corporation, Ankara, Turkey.<br>- [b]Department of Cardiology, Amsterdam Cardiovascular Sciences, Amsterdam University Medical Centre, University of Amsterdam, Amsterdam, Netherlands<br>- [c]Institute of Health Informatics, University College London, London, UK<br>- [d]The National Institute for Health Research University College London Hospitals Biomedical Research Centre, University College London, London, UK<br>- [e]Artificial Intelligence in Medicine Lab (BCN-AIM), Departament de Matemàtiques i Informàtica, Universitat de Barcelona, Barcelona, Spain<br>- [f]Institució Catalana de Recerca i Estudis Avançats (ICREA), Barcelona, Spain |
| **Corresponding Author** | A. Anil SINACI, PhD.<br>Principal Researcher.<br><br>Address: SRDC A.S. ODTU Teknokent Silikon Bina K1-16 06800 Cankaya, Ankara, Turkiye.<br>E-Mail: anil@srdc.com.tr<br>Tel: +90 312 2101763<br>Fax: +90 312 2101837 |
| **Keywords** | AI Provenance; Healthcare AI; Transparency; Traceability; AI Documentation; Digital Passports |
| **Word count** | Abstract: 243<br>Main text: 3002 |

# Abstract

**Objective:** To develop the AI Product Passport, a standards-based framework enhancing transparency, traceability, and regulatory compliance in healthcare AI by providing lifecycle-based documentation.

**Materials and Methods:** The AI Product Passport was developed within the AI4HF project, focusing on heart failure AI tools. We analyzed regulatory frameworks (EU AI Act, FDA guidelines) and existing standards (PROV-ML, Model Cards, FactSheets) to design a relational data model capturing metadata across AI lifecycle phases: study definition, dataset preparation, model generation/evaluation, deployment/monitoring, and passport generation. MLOps/ModelOps concepts were integrated for operational relevance. Co-creation involved feedback from AI4HF consortium members and a Lisbon workshop with 21 diverse stakeholders, evaluated via Mentimeter polls. The open-source platform was implemented with Python libraries for automated provenance tracking.

**Results:** The AI Product Passport was designed based on existing standards and methods with well-defined lifecycle management and role-based access. Its implementation is a web-based platform with a relational data model supporting comprehensive, auditable documentation. It generates machine-readable metadata and human-readable reports, customizable for stakeholders (clinicians, developers, regulators). It aligns with FUTURE-AI principles (Fairness, Universality, Traceability, Usability, Robustness, Explainability), ensuring fairness, traceability, and usability. Exported passports detail model purpose, data provenance, performance, and deployment context. GitHub-hosted backend/frontend codebases enhance accessibility.

**Discussion and Conclusion:** The AI Product Passport addresses transparency gaps in healthcare AI, meeting regulatory and ethical demands. Its open-source nature and alignment with standards foster trust and adaptability. Future enhancements include FAIR data principles and FHIR integration for improved interoperability, promoting responsible AI deployment.

## 1. Introduction

Artificial Intelligence (AI) is increasingly transforming healthcare, supporting predictive analytics, diagnostics, and personalized medicine. As AI systems become more embedded in clinical workflows, ensuring their reliability, fairness, and transparency has become critical to maintaining public trust and enabling safe, ethical deployment. Healthcare AI faces challenges in transparency, reproducibility, and equitable usability, limiting clinical trust. In response, global regulatory bodies are developing stricter oversight frameworks. The European Union's AI Act and U.S. FDA guidance on Software as a Medical Device emphasize the importance of transparency, traceability, and accountability. Yet current regulations provide little operational guidance, leaving developers without practical tools to meet documentation and monitoring expectations across the full AI lifecycle.

To address this gap, we propose the AI Product Passport: a structured framework that enables comprehensive, lifecycle-based documentation of healthcare AI systems. The AI Product Passport integrates established standards into a unified, modular system. It supports the generation of both machine-readable and human-readable outputs to serve the needs of diverse stakeholders including clinicians, developers, and regulators. The design of the AI Product Passport is grounded in the FUTURE-AI guidelines[1], which define core principles for trustworthy AI in healthcare: Fairness, Universality, Traceability, Usability, Robustness, and Explainability. These principles inform both the technical architecture and the ethical underpinnings of our framework. In this study, we describe the rationale, development, and structure of the AI Product Passport, and demonstrate how it supports regulatory compliance, ethical transparency, and operational usability in real-world healthcare settings. Our goal is to provide a practical and open-source solution that fosters responsible AI development and deployment across the healthcare ecosystem.

## 2. Background

The ability to manage and document the origins, changes, and usage of data, known as provenance management[2], has become increasingly important in AI[3]. Effective provenance supports reproducibility and regulatory compliance by documenting how models are trained, tested, and applied. However, regulatory frameworks like the EU AI Act[4] and U.S. FDA guidelines[5], while emphasizing transparency, provide limited practical guidance. The EU AI Act mandates transparency reports for high-risk systems but lacks specifics on documenting decision-making. Similarly, FDA guidance highlights real-world monitoring and explainability without structured methods for tracing model changes or training data.

Existing frameworks such as PROV-ML[6] (for data lineage) which is an extension of the W3C PROV[7] standard, IBM FactSheets[8] (for structured model reports), and Model Cards[9] (for concise model summaries) address AI transparency but each covers limited lifecycle phases. In addition to documentation-focused approaches, MLOps[10] and ModelOps[11] apply DevOps principles to manage AI workflows and governance, enabling version control, monitoring, and compliance throughout model lifecycles.

Complementing these technical and operational approaches, domain-specific frameworks have also been introduced to guide the development of trustworthy AI in healthcare. One prominent example is the FUTURE-AI guidelines[1], a consensus-driven set of principles for medical AI. Rather than focusing solely on technical metrics like accuracy or performance, these principles emphasize broader responsibilities, such as ethical alignment, usability for diverse stakeholders, and transparent communication across the AI lifecycle.

While the abovementioned technical frameworks provide valuable insights, they still have limitations when applied to healthcare AI. Many focus only on specific phases of the AI lifecycle, such as model evaluation, and do not fully address the need for continuous, end-to-end traceability required in regulated clinical environments. Furthermore, the integration of ethical principles, stakeholder usability, and explainability is limited in practice. Persistent challenges exist in applying provenance mechanisms effectively to healthcare applications. First, data used to train AI models often originates from heterogeneous sources, including electronic health records, registries, medical imaging, and wearable technologies, making it difficult to reconstruct a complete data history[12]. Second, more advanced healthcare AI models are typically complex and not inherently interpretable, complicating efforts by clinicians and regulators to understand and evaluate predictions[13]. While some healthcare models, such as MAGGIC[14], are highly transparent and interpretable, others – particularly deep learning models used for tasks like segmentation or risk prediction – tend to lack such explainability. Third, although existing standards provide a foundation for capturing provenance, their implementation in operational workflows remains inconsistent and fragmented. Institutions frequently adopt varying documentation practices, which makes it difficult to compare, audit, or integrate AI models across clinical settings[15]. These gaps highlight the need for comprehensive, standardized, end-to-end and context-aware approaches for AI documentation in healthcare. In response to this need, we have designed and developed the AI Product Passport following standardized methodologies in alignment with FUTURE-AI principles and released as permissive open-source projects.

## 3. Methods

The development of the AI Product Passport was initiated within the scope of the AI4HF project[16], which aims to develop AI-based tools to support heart failure prevention, diagnosis, and management by leveraging trustworthy and explainable AI methods. Within this broader objective, the AI Product Passport aims to enhance transparency, traceability, and regulatory compliance in AI-driven healthcare solutions. Design and development methodology was characterized by an iterative process as illustrated in Figure 1.

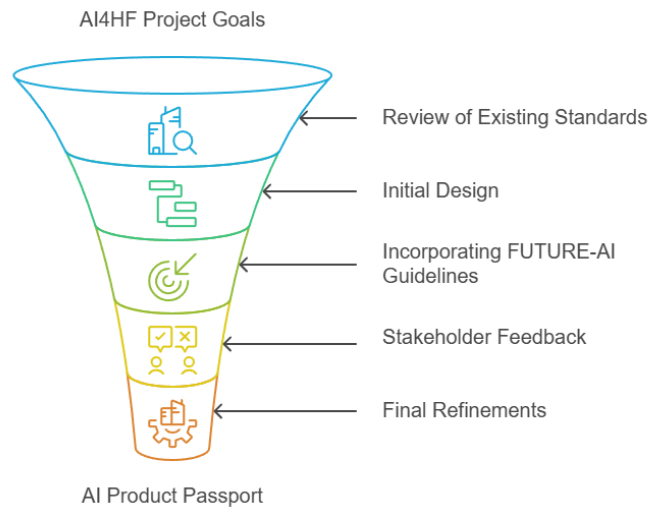

*Figure 1: AI Product Passport development process*

The documentation challenges were identified, and key regulations and standards were analyzed to design a relational data model combining elements of PROV-ML, Model Cards, and FactSheets for traceable, auditable reporting. A mapping exercise was carried out by consortium members to identify and select relevant metadata entities from these standards. Initially, 38 candidate metadata entities were compiled. Following two rounds of review and discussion, this set was refined to 32 entities, which formed the foundation of the first version of the data model.

The first version incorporated provenance tracking to document AI models across their lifecycle. To enhance the operational relevance and adaptability of the framework, the initial data model was extended by incorporating concepts from MLOps and ModelOps, such as version control and deployment-specific metadata. These additions were critical for supporting real-world use cases, where AI models are subject to continuous updates, performance monitoring, and compliance-related documentation throughout their lifecycle. Building on these foundations, the system was designed to align not only with provenance standards but also with best practices in model governance and operational integration.

In addition to aligning with technical standards, the development process was guided by the principles defined in the FUTURE-AI guidelines. Rather than being treated as abstract ideals, the six FUTURE-AI principles were used to inform practical questions throughout the design process; for example, how to structure documentation in a way that supports clarity, how to make the system usable across stakeholder groups and how to promote reproducibility and accountability in model development. This alignment influenced the definition of both functional requirements and reporting mechanisms during the initial and iterative stages of the development of the framework.

A co-creation process involved AI4HF consortium members and 21 Lisbon workshop participants representing clinical, technical, and patient perspectives. An interactive assessment session was held during the Lisbon workshop, where participants responded to a series of live questions using Mentimeter, an online tool for real-time polling and feedback. These structured polls captured stakeholder preferences regarding usability and robustness. Figure 2 shows an example of one of the poll questions and the participant responses. The insights gathered from this session contributed directly to the refinement of the AI Product Passport, particularly in simplifying documentation outputs and enhancing reporting flexibility. Throughout the project, continuous feedback from multidisciplinary stakeholders guided successive refinements, allowing the AI Product Passport framework to evolve in response to both technical demands and practical requirements.

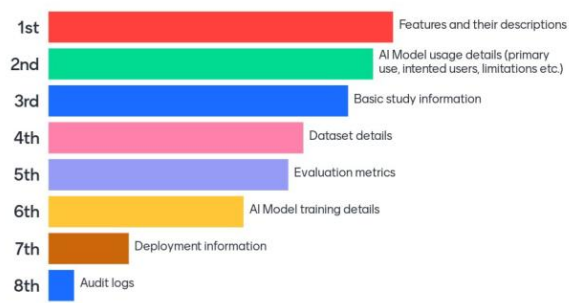
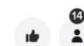

*Figure 2: An example Mentimeter poll from the interactive session in Lisbon Workshop*

## 4. Results

The AI Product Passport was developed as a web-based platform to facilitate structured, transparent, and traceable documentation across the entire lifecycle of AI models used in healthcare. Its development is open source; backend and frontend code bases are maintained as separate GitHub projects[17,18]. At the core of the platform is a structured, relational data model grounded on well-established standards, specifically designed to capture key metadata, decisions, and configurations made during the development, evaluation, and deployment of AI models. This data model enables systematic documentation of both technical processes and contextual elements, including model purpose, data provenance, evaluation results, and deployment conditions. It also supports consistency and reproducibility by organizing this information into modular and logically linked components aligned with each phase of the AI lifecycle.

### 4.1 Lifecycle Management

The relational data model developed for the AI Product Passport supports comprehensive documentation across five distinct phases of the AI model lifecycle as depicted in Figure 3. Designed to accommodate the needs of diverse stakeholders, the model allows both technical detail and narrative-level summaries to coexist. It supports the generation of machine-readable metadata and human-readable documentation, offering flexibility for integration with various institutional workflows. Each phase captures key aspects of provenance and model governance, enabling consistent traceability, auditability, and transparency across the study, development, and deployment of healthcare AI systems.

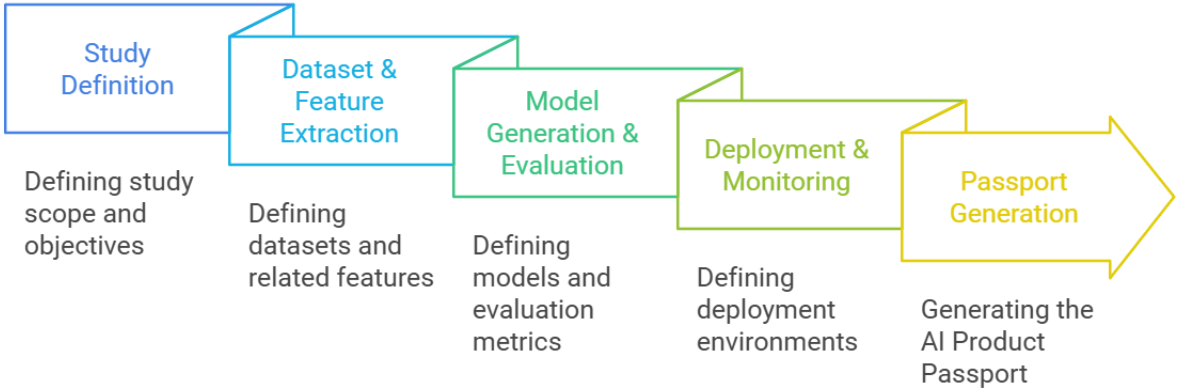

*Figure 3: AI Product Passport documentation lifecycle*

### 4.1.1 Phase 1: Study Definition

This initial phase involves defining the scope and objectives of the AI study, including organizational parameters and data collection strategies that guide subsequent stages of model development. Elements from FactSheets are used to document the intended use of the AI system, the clinical or regulatory context in which it operates, and relevant ethical considerations. Preliminary surveys or study configurations such as study goals, population characteristics, and research questions are also captured as foundational data. All study setup information is recorded in the AI Product Passport to establish a clear and traceable point of origin.

### 4.1.2 Phase 2: Dataset & Featureset Extraction

In this phase, metadata about the datasets used for model training and validation is collected and documented. This includes information on data sources, the methods applied to clean and transform the data, and how the datasets were split into training, validation, and test sets. PROV-ML constructs are used to capture the provenance of these operations, enabling traceability of the data preparation pipeline. Metadata on features, datasets, and transformations are structured as shown in Table 1.

*Table 1: Data model entities used in Phase 1 & Phase 2*

|  | Entity | Description |
|---|---|---|
| **Phase 1** | Organization | The institution, department, or entity responsible for initiating, overseeing, or participating in the clinical study or project. |
|  | Personnel | Individuals from an organization contributing to the study, including roles in design, development, testing, or management. |
|  | Study | The overarching investigation, defined by a research hypothesis or objective, guiding the AI/ML workflow. |
|  | Experiment | A specific set of analyses or research questions within a Study, each targeting a distinct outcome variable for AI/ML modeling. For a predictive model, each outcome variable will correspond to a different research question. |
|  | Survey | Structured textual questions and responses capturing ethical approvals, data governance policies, identified biases, and study limitations. |

|  | Feature | A single column or attribute in a tabular, AI-ready dataset, representing a measurable variable. |
|---|---|---|
| **Phase 2** | Featureset | A logically grouped collection of Features used to construct a Dataset for a specific Experiment. |
|  | Dataset | A structured collection of data, shaped according to a Featureset, containing Feature values for a defined Experiment. |
|  | LearningDataset | The final dataset, derived from a Dataset after applying transformations (e.g., dimensionality reduction, imputation, normalization), used as direct input to AI/ML algorithms. |
|  | DatasetTransformation | The sequence of data processing steps (e.g., cleaning, scaling, encoding) applied to transform a Dataset into a LearningDataset. |

### 4.1.3 Phase 3: Model Generation & Evaluation

This phase focuses on documenting how AI models are developed, trained, and evaluated using the datasets prepared in Phase 2. The AI Product Passport captures metadata related to the modeling process, including the algorithms and libraries used, training parameters, and performance evaluation metrics. From Model Cards, the framework adopts fields for describing intended use, performance metrics, and ethical considerations related to model evaluation. PROV-ML is used to structure provenance data related to model training steps, while elements from Model Cards guide the documentation of evaluation metrics, validation procedures, and performance results. Table 2 lists the entities and their descriptions modeled to cover this phase.

*Table 2: Data model entities used in Phase 3 & Phase 4*

|  | **Entity** | **Description** |
|---|---|---|
| **Phase 3** | Algorithm | A machine learning technique (e.g., k-means clustering, random forest) defined independently of its software implementation. |
|  | Implementation | The specific software or library (e.g., scikit-learn, TensorFlow) that realizes an Algorithm for practical use. |
|  | LearningProcess | The execution of an AI/ML workflow on a LearningDataset using a specific Implementation to train or evaluate a model. |
|  | LearningStage | A distinct phase in the LearningProcess (e.g., training, testing, validation, or federated model aggregation). |
|  | Model | The trained machine learning model resulting from a LearningProcess, ready for evaluation or deployment. |
|  | Parameter | Configurable settings or hyperparameters used in an Algorithm, LearningProcess, LearningStage, or Model to optimize performance. |
|  | EvaluationMeasure | A metric (e.g., accuracy, sensitivity, F1-score) used to assess the performance of a Model during evaluation. |
| **Phase 4** | ModelDeployment | The process of deploying a trained Model to a specific operational environment for real-world use. |
|  | DeploymentEnvironment | The hardware and software configuration (e.g., cloud, on-premises server, edge device) hosting a ModelDeployment. |

### 4.1.4 Phase 4: Deployment & Monitoring

This phase focuses on documenting information related to the deployment of the models. The AI Product Passport records key operational metadata, including the deployment context, environment specifications, and monitoring protocols. Inspired by FactSheets and ModelOps principles, this documentation provides a clear view of how and where the model is intended to function in real-world settings. Table 2 describes the two entities used in Phase 4; ModelDeployment and DeploymentEnvironment.

### 4.1.5 Phase 5: Passport Generation

In the final phase, all accumulated metadata and documentation outputs from the previous stages are compiled into a structured AI Product Passport. This document is designed to serve multiple audiences and can be configured to present varying levels of detail. The final passport includes comprehensive records of conceptualization, development, evaluation, and deployment of the AI model (Table 3). It is formatted for export in a standardized layout and can be secured using Digital Signature Service (DSS)[19] to maintain integrity and authenticity.

*Table 3: Data model entities used in Phase 5*

| Entity | Description |
| --- | --- |
| Passport | The AI Product Passport, a comprehensive document aggregating machine-readable metadata and human-readable documentation from all lifecycle phases. |
| AuditLog | A record of actions (e.g., creation, update, deletion) performed on data entities, including timestamps and responsible personnel. |
| AuditLogBook | A consolidated collection of AuditLogs associated with a specific Passport instance, ensuring traceability and auditability. |

The complete data model is published as an Entity-Relationship diagram along with SQL Data Definition Language (DDL) scripts in the Git repository[17].

## 4.2 Role-based Management

In operational settings, the AI Product Passport is intended to be collaboratively maintained. To enable this collaboration, users are assigned different roles, each with specific responsibilities as described in Table 4.

*Table 4: Role-based management of an AI Product Passport throughout its lifecycle*

| Role Name | Description | Phase |
| --- | --- | --- |
| Study Owner | Defines the study, assigns personnel and their roles, specifies the study population, and formulates research questions. | Phase 1: Study Definition |
| Survey Manager | Defines survey questions and answers related to the study. | Phase 1: Study Definition |
| Data Engineer | Defines feature sets, related features, outcomes, and dataset information such as characteristics, transformations, and metadata. | Phase 2: Dataset & Feature Set Extraction |
| Data Scientist | Defines model parameters and metadata, links datasets with models, and creates learning process objects. | Phase 3: Model Generation & Evaluation |

| ML Engineer | Defines deployment-related information, including hardware and software environments in which the model operates. | Phase 4: Deployment & Monitoring |
|---|---|---|
| Quality Assurance Specialist | Creates and maintains model passports, manages versions, and configures the level of detail for exported passports. | Phase 5: Passport Generation |

### 4.3 Alignment with FUTURE-AI

In alignment with broader efforts to define trustworthy AI in healthcare, the framework also reflects the six principles established in FUTURE-AI guidelines. These principles provided a conceptual foundation for translating ethical values into practical features throughout the design and implementation of the framework. Table 5 summarizes how the AI Product Passport responds to each of the six principles defined in FUTURE-AI, demonstrating how technical implementation choices support ethical and stakeholder-aligned outcomes.

*Table 5: Implementation of FUTURE-AI Principles in AI Product Passport*

| FUTURE-AI principle | How AI Product Passport aligns with the principle |
|---|---|
| Fairness | Supported through transparent reporting mechanisms that help reveal potential biases in training data, feature selection, and model performance outcomes. |
| Universality | Addressed by offering flexible export functionality that allows users to select the level of detail appropriate to their background and purpose. |
| Traceability | Achieved through structured metadata capture across all phases of model development, including study definition, model training, performance evaluation and deployment. |
| Usability | Enabled by a simplified AI product passport export interface and clear lifecycle structuring, allowing users with different levels of technical expertise to navigate the documentation. |
| Robustness | Addressed by documenting model training conditions, data characteristics, and evaluation results, which helps identify potential sources of variation |
| Explainability | Addressed by clearly documenting model inputs, parameters, and outcome metrics to help stakeholders interpret how models produce their results. |

### 4.4 Product Passports

The AI Product Passport allows users to export AI documentation as PDF files, providing flexibility in specifying the level of detail based on predefined categories. Users can select specific aspects of the lifecycle of the AI model, such as data preprocessing steps, model training details, evaluation metrics, and deployment records, making the documentation more relevant to their intended audience. This structured reporting approach supports a broad range of stakeholders, including regulators who require compliance records, developers who need technical insights, and healthcare professionals who seek clear explanations of model behavior. By offering customizable documentation, the AI Product Passport facilitates more efficient auditing, informed decision-making, and better communication among different stakeholders in the AI ecosystem.

To illustrate how these features come together in practice, an example of a generated AI Product Passport is presented in Figure 4. The passport presents a structured summary of key information from the AI lifecycle in a format that is accessible and easy to interpret. It includes essential metadata such as model identity, deployment context, computational environment, datasets used, and detailed feature information. The level of detail included in each generated passport can be configured by the user at the time of export, allowing the document to be adapted to different informational needs. This flexibility makes the passport suitable for a wide range of stakeholders, including clinicians, developers, auditors, and regulators. In practice, the passport serves as a bridge between technical processes and stakeholder understanding by translating lifecycle metadata into a clear, navigable format. This enhances transparency in the use of AI tools and helps stakeholders evaluate models with greater clarity and confidence.

A step-by-step guide to using the web-based system for generating an AI Product Passport is provided in the Supplementary Material. It also demonstrates how to submit passport data directly from within Python code during model training using the associated Python library.

**AI4HF Product Passport**

## Model Details

| | |
|---|---|
| Model Name | MAGGIC-MLP Model (v1.0) |
| Model Version | 1.0 |
| Model Type | Classification |
| Product Identifier | AI4HF_MAGGIC_MLP_001 |
| Owner | Amsterdam UMC |
| TRL Level | TRL6 |
| License | AI4HF-Research License v1.0 |
| Primary Use | Predicting 1-year mortality risk in chronic heart failure patients based on MAGGIC feature set. |
| Secondary Use | Clinical risk stratification and care prioritization. |
| Intended Users | Clinicians, cardiologists, and data science researchers. |
| Counter Indications | Not suitable for pediatric Chronic Heart Failure or congenital heart disease populations. |
| Ethical Considerations | Ethically approved for retrospective analysis; patient identifiers are anonymized. |
| Limitations | Performance may degrade in cohorts with limited lab or echocardiographic data. |
| Fairness Constraints | Bias mitigation methods were applied across gender and age subgroups. |

## Model Deployment Details

| | |
|---|---|
| Deployment Tags | Validation, ProductionCandidate |
| Deployment Status | VALIDATING |
| Identified Failures | Model occasionally overestimates low-risk cases with missing LVEF values. |

## Environment Details

| | |
|---|---|
| Environment Title | Clinical Validation Environment |
| Environment Description | Dedicated environment for MAGGIC-MLP model validation under clinical conditions. |
| Hardware Properties | RAM: 64 GB, CPU: 16 cores, GPU: 1x NVIDIA A100 40GB |
| Software Properties | OS: Ubuntu 22.04, Frameworks: TensorFlow 2.15, PyTorch 2.2, Spark 3.5 |
| Connectivity Details | Secure hospital intranet connection with role-based VPN access and encrypted endpoints. |

## Feature Sets

| FeatureSet Title | Feature Set Description | FeatureSet URL | Created At | Created By | Last Updated At | Last Updated By |
|---|---|---|---|---|---|---|
| MAGGIC Score Predictors | Canonical predictors used in MAGGIC score calculation and model training. | https://ai4hf.eu/feature-sets/maggic-v1 | Oct 15, 2025, 12:00:00 AM | data_engineer | Oct 15, 2025, 12:00:00 AM | data_engineer |

| Feature Title | Feature Description | Feature Data Type | Is an Outcome | Feature Mandatory | Feature Unique | Feature Units | Feature Equipment | Feature Data Collection |
|---|---|---|---|---|---|---|---|---|
| gender | Gender of the patient | string | No | Yes | No | | | EHR |
| age | Age of the patient at reference point | integer | No | Yes | No | years | | EHR |
| nyha | The latest value of the New York Heart Assessment as LOINC Code | string | No | Yes | No | | | EHR/Clinical assessment |
| systolic_blood_pressure | Average systolic blood pressure (mmHg) over 3 years preceding | decimal | No | Yes | No | mmHg | | Vitals |

*Figure 4: An example of generated AI product passport for an AI model calculating Heart Failure Risk Score*

## 5. Discussion

The development of the AI Product Passport aligns closely with broader policy initiatives on AI transparency, particularly the recommendations outlined in the Policy Alignment on AI Transparency[20] report by the Partnership on AI. This report emphasizes the need for structured and standardized documentation across the AI lifecycle to improve accountability and trust. The AI Product Passport reflects these principles through lifecycle-based documentation capturing study definition, data processing, model development, evaluation, and deployment. Its integration of recognized standards demonstrates a commitment to transparency and traceability at every stage. The report also highlights the importance of supplier declarations of conformity, encouraging verifiable, comprehensive AI disclosures. The AI Product Passport addresses this recommendation by consolidating detailed provenance records and structured summaries of model behavior into a standardized, exportable format that can be shared with regulators, healthcare providers, and other stakeholders. Moreover, the report emphasizes the need for transparency mechanisms to be accessible to a diverse range of users, including non-technical stakeholders. The AI Product Passport supports this by allowing users to customize the detail level, ensuring clarity for each audience. Through its emphasis on end-to-end documentation, integration of established standards, and support for customizable reporting, the AI Product Passport implements several core principles from the Policy Alignment on AI Transparency report, contributing to responsible and transparent AI deployment in healthcare.

Existing frameworks previously mentioned such as IBM Factsheets, Model Cards and PROV-ML have contributed significantly to advancing transparency, fairness, and provenance management in AI systems. However, these frameworks typically address specific components of AI documentation rather than offering an end-to-end, domain-specific solution. FactSheets and Model Cards focus primarily on summarizing model characteristics, performance, intended use, and ethical considerations, which are valuable but often lack deeper integration with the operational workflows and continuous lifecycle management needed in regulated sectors like healthcare. Similarly, PROV-ML provides a strong foundation for capturing provenance within machine learning workflows but does not directly offer user-friendly, role-based reporting mechanisms suited for healthcare professionals, developers, and regulators.

The AI Product Passport unifies and extends these frameworks into a single platform covering the entire AI lifecycle – from study design to deployment – adding structured, exportable reporting tailored for regulators and clinicians. This integrated approach bridges the gaps left by existing frameworks, providing both technical depth and practical usability within a highly regulated domain, while also reflecting the broader ethical and stakeholder-oriented expectations emphasized in the FUTURE-AI guidelines.

Future work will enhance the interoperability and openness of AI Product Passport instances by adopting FAIR (Findable, Accessible, Interoperable, Reusable) principles[21]. Passport metadata will be published through FAIR Data Point mechanisms[22] using DCAT-based formats[23] and linked to persistent identifiers such as DOIs. Following initiatives such as FAIR for AI[24–26] and the Gardens framework[27], this approach will expose passport information via RESTful APIs, enabling both human- and machine-readable access to models, datasets, validation results, and limitations. Aligning with FAIR standards will improve discoverability, interoperability, and reusability of healthcare-AI documentation, fostering transparent and collaborative reuse across institutions. In parallel, we plan to define a machine-readable DatasetTransformation entity aligned with declarative pipelines for AI-ready data[28] and HL7 FHIR standards[29]. Integrating this into the Passport will capture preprocessing steps from raw data to learning datasets, further strengthening transparency and traceability in healthcare-AI development.

## 6. Conclusion

The AI Product Passport establishes a practical foundation for transparent and accountable healthcare AI. By unifying standards such as PROV-ML, FactSheets, Model Cards, and MLOps/ModelOps, it delivers auditable documentation that fulfills regulatory and ethical expectations under the EU AI Act and FDA guidelines. Its alignment with FUTURE-AI principles ensures usability and trust across stakeholders. Released as open source, the framework invites community adoption, while planned FAIR and FHIR integrations will expand its interoperability and long-term impact.

## 7. Acknowledgements

This work was supported by the AI4HF project[16], which has received funding from the European Union's Horizon Research and Innovation Programme under Grant Agreement No 101080430.

## 8. Declaration of AI-assisted technologies

The authors acknowledge the use of ChatGPT (OpenAI) for assistance with English language editing and rephrasing during manuscript preparation. All outputs were carefully reviewed and edited by the authors, who assume full responsibility for the content of the published article.

## 9. References

[1] Lekadir K, Frangi AF, Porras AR, Glocker B, Cintas C, Langlotz CP, et al. FUTURE-AI: international consensus guideline for trustworthy and deployable artificial intelligence in healthcare. BMJ 2025:e081554. https://doi.org/10.1136/bmj-2024-081554.

[2] Herschel M, Diestelkämper R, Ben Lahmar H. A survey on provenance: What for? What form? What from? The VLDB Journal 2017;26:881–906. https://doi.org/10.1007/s00778-017-0486-1.

[3] Souza R, Azevedo L, Lourenco V, Soares E, Thiago R, Brandao R, et al. Provenance Data in the Machine Learning Lifecycle in Computational Science and Engineering. 2019 IEEE/ACM Workflows in Support of Large-Scale Science (WORKS), IEEE; 2019, p. 1–10. https://doi.org/10.1109/WORKS49585.2019.00006.

[4] European Parliament and Council of the European Union. Regulation (EU) 2024/1689 of the European Parliament and of the Council of 13 June 2024 laying down harmonised rules on artificial intelligence and amending Regulations (EC) No 300/2008, (EU) No 167/2013, (EU) No 168/2013, (EU) 2018/858, (EU) 2018/1139 and (EU) 2019/2144 and Directives 2014/90/EU, (EU) 2016/797 and (EU) 2020/1828 (Artificial Intelligence Act). http://data.europa.eu/eli/reg/2024/1689/oj; 2024.

[5] U.S. Food and Drug Administration. Artificial Intelligence and Machine Learning in Software as a Medical Device | FDA. 2025.

[6] Souza R, Azevedo LG, Lourenço V, Soares E, Thiago R, Brandão R, et al. Workflow provenance in the lifecycle of scientific machine learning. Concurr Comput 2022;34:e6544. https://doi.org/10.1002/CPE.6544;REQUESTEDJOURNAL:JOURNAL:15320634;WGROUP:STRING:PUBLICATION.

[7] Missier P, Belhajjame K, Cheney J. The W3C PROV family of specifications for modelling provenance metadata. ACM International Conference Proceeding Series 2013:773–6. https://doi.org/10.1145/2452376.2452478;SUBPAGE:STRING:ABSTRACT;CSUBTYPE:STRING:CONFERENCE.

[8] Arnold M, Piorkowski D, Reimer D, Richards J, Tsay J, Varshney KR, et al. FactSheets: Increasing trust in AI services through supplier's declarations of conformity. IBM J Res Dev 2019;63. https://doi.org/10.1147/JRD.2019.2942288.

[9] Mitchell M, Wu S, Zaldivar A, Barnes P, Vasserman L, Hutchinson B, et al. Model cards for model reporting. FAT* 2019 - Proceedings of the 2019 Conference on Fairness, Accountability, and Transparency 2019:220–9. https://doi.org/10.1145/3287560.3287596;PAGE:STRING:ARTICLE/CHAPTER.

[10] Kreuzberger D, Kuhl N, Hirschl S. Machine Learning Operations (MLOps): Overview, Definition, and Architecture. IEEE Access 2023;11:31866–79. https://doi.org/10.1109/ACCESS.2023.3262138.

[11] Hummer W, Muthusamy V, Rausch T, Dube P, El Maghraoui K, Murthi A, et al. ModelOps: Cloud-based lifecycle management for reliable and trusted AI. Proceedings - 2019 IEEE International Conference on Cloud Engineering, IC2E 2019 2019:113–20. https://doi.org/10.1109/IC2E.2019.00025.

[12] Holzinger A, Keiblinger K, Holub P, Zatloukal K, Müller H. AI for life: Trends in artificial intelligence for biotechnology. N Biotechnol 2023;74:16–24. https://doi.org/10.1016/J.NBT.2023.02.001.

[13] Amann J, Blasimme A, Vayena E, Frey D, Madai VI. Explainability for artificial intelligence in healthcare: a multidisciplinary perspective. BMC Med Inform Decis Mak 2020;20:1–9. https://doi.org/10.1186/S12911-020-01332-6/PEER-REVIEW.

[14] Pocock SJ, Ariti CA, McMurray JJV, Maggioni A, Køber L, Squire IB, et al. Predicting survival in heart failure: a risk score based on 39 372 patients from 30 studies. Eur Heart J 2013;34:1404–13. https://doi.org/10.1093/eurheartj/ehs337.

[15] Khoi Tran N, Sabir B, Babar MA, Cui N, Abolhasan M, Lipman J. ProML: A Decentralised Platform for Provenance Management of Machine Learning Software Systems. Lecture Notes in Computer Science (Including Subseries Lecture Notes in Artificial Intelligence and Lecture Notes in Bioinformatics) 2022;13444 LNCS:49–65. https://doi.org/10.1007/978-3-031-16697-6_4.

[16] AI4HF Project Consortium. Trustworthy Artificial Intelligence for Personalised Risk Assessment in Chronic Heart Failure (AI4HF Project) n.d. https://www.ai4hf.com/ (accessed May 21, 2025).

[17] AI4HF/passport: The backend of AI4HF Product Passport n.d. https://github.com/AI4HF/passport (accessed June 24, 2025).

[18] AI4HF/passport-web: The user interface of AI4HF Product Passport n.d. https://github.com/AI4HF/passport-web (accessed June 24, 2025).

[19] European Commission. Digital Signature Service 2024. https://ec.europa.eu/digital-building-blocks/DSS/webapp-demo/doc/dss-documentation.html (accessed May 27, 2025).

[20] Howell J, Ifayemi S. PARTNERSHIP ON AI Policy Alignment on AI Transparency Policy Alignment on AI Transparency Analyzing Interoperability of Documentation Requirements across Eight Frameworks n.d.

[21] Wilkinson MD, Dumontier M, Aalbersberg IjJ, Appleton G, Axton M, Baak A, et al. The FAIR Guiding Principles for scientific data management and stewardship. Sci Data 2016;3:1–9. https://doi.org/10.1038/SDATA.2016.18;SUBJMETA=479,648,697,702,706;KWRD=PUBLICATION+CHARACTERISTICS,RESEARCH+DATA.

[22] da Silva Santos LOB, Burger K, Kaliyaperumal R, Wilkinson MD. FAIR Data Point: A FAIR-Oriented Approach for Metadata Publication. Data Intell 2025;5:163–83. https://doi.org/10.1162/DINT_A_00160.

[23] W3C. Data Catalog Vocabulary (DCAT) - Version 3 2024. https://www.w3.org/TR/vocab-dcat-3/ (accessed May 30, 2025).

[24] Huerta EA, Blaiszik B, Brinson LC, Bouchard KE, Diaz D, Doglioni C, et al. FAIR for AI: An interdisciplinary and international community building perspective. Sci Data 2023;10:1–10. https://doi.org/10.1038/S41597-023-02298-6;SUBJMETA=496,648,697,706;KWRD=RESEARCH+DATA,RESEARCH+MANAGEMENT.

[25] Duarte J, Li H, Roy A, Zhu R, Huerta EA, Diaz D, et al. FAIR AI models in high energy physics. Mach Learn Sci Technol 2023;4:045062. https://doi.org/10.1088/2632-2153/ad12e3.

[26] Ravi N, Chaturvedi P, Huerta EA, Liu Z, Chard R, Scourtas A, et al. FAIR principles for AI models with a practical application for accelerated high energy diffraction microscopy. Sci Data 2022;9:657. https://doi.org/10.1038/s41597-022-01712-9.

[27] Engler W. Frameworks: Garden: A FAIR framework for publishing and applying AI models for translational research in science, engineering, education, and industry. PI: Ian Foster. Co-PIs: Ben Blaiszik, Eliu Huerta, Dane Morgan, Rebecca Willett, Rafael Gomez-Bombarelli. 2023. https://doi.org/10.6084/m9.figshare.24212829.v1.

[28] Namli T, Anıl Sınacı A, Gönül S, Herguido CR, Garcia-Canadilla P, Muñoz AM, et al. A scalable and transparent data pipeline for AI-enabled health data ecosystems. Front Med (Lausanne) 2024;11:1393123. https://doi.org/10.3389/FMED.2024.1393123/BIBTEX.

[29] Health Level 7 (HL7). Fast Healthcare Interoperability Resources (FHIR) n.d. https://hl7.org/fhir/ (accessed May 21, 2025).

# Supplementary: User Guide for the AI Product Passport Web Interface

## 1. Introduction

This supplementary material provides a visual walkthrough of the application's user interface, complementing the descriptions presented in the main manuscript. Each page of the interface is illustrated through screenshots, accompanied by brief explanations highlighting key functionalities and design choices. The aim is to offer readers a clearer understanding of the user experience and how the system supports the workflows and objectives discussed in the study.

## 2. Login

The interface design of the AI Product Passport tool focuses on ease of use and effective interaction for users across different roles. Once a user logs in through a clean and straightforward interface (Figure 1), the application dynamically configures itself based on the user's assigned role, presenting only the relevant functionalities and interfaces tailored to their specific responsibilities. On the backend, an authorization server supports this process by integrating an identity and access management that enables secure access to the system while simplifying the management of user roles and permissions.

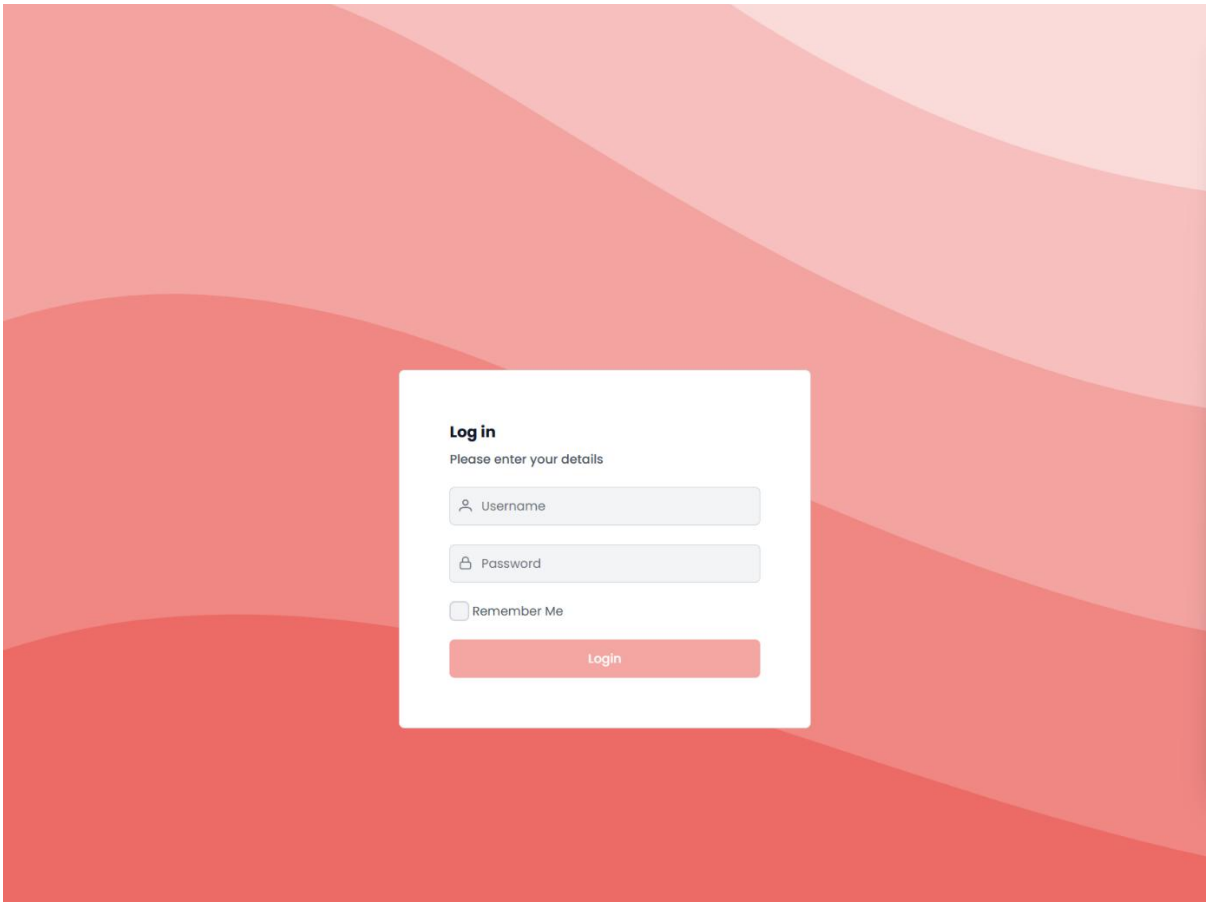

*Figure 1: Login Interface*

Available user roles defined in AI Product Passport are as follows:

1. **Admin:** Responsible for overall system administration including user & identity management, system-level configurations, security updates and maintenance.
2. **Organization Admin:** Manages organizational profiles, including the creation of new organizational entities and the addition of personnel.
3. **Study Owner:** Manages studies, including the setup, coordination, and closure of study activities. This role also involves inviting and managing participants' roles within specific studies.
4. **Survey Manager:** Handles the creation, deployment, and analysis of surveys that gather critical information related to ongoing studies, playing a key role in data collection processes.
5. **Data Engineer:** Focuses on managing datasets and performing feature extraction tasks, ensuring that data is processed and prepared accurately for further analysis or model training.
6. **Data Scientist:** Leads model generation and validation efforts, applying scientific methods and algorithms to develop and refine predictive models based on the prepared datasets.
7. **ML Engineer:** Handles the deployment of ML models into production environments, managing the technical aspects of deploying and scaling AI solutions.
8. **Quality Assurance Specialist:** Ensures that all components of the AI Product Passport meet the required quality standards, including validating and approving the passports generated for compliance and accuracy.

## 3. Study Management

The Study Management interface is designed to support the responsibilities of the Study Owner, who oversees the setup, coordination, and closure of study activities (Figure 2). Through this interface, the Study Owner can define and manage studies related to AI workflows by specifying key details such as study objectives, population criteria, personnel assignments, and experiment design (Figure 3). Additionally, the interface provides tools for inviting participants and managing their roles within each study, streamlining the overall study coordination process. Each role, except for the study owner, has access to a study selection interface. They can choose a study to work on and have permission to view its details.

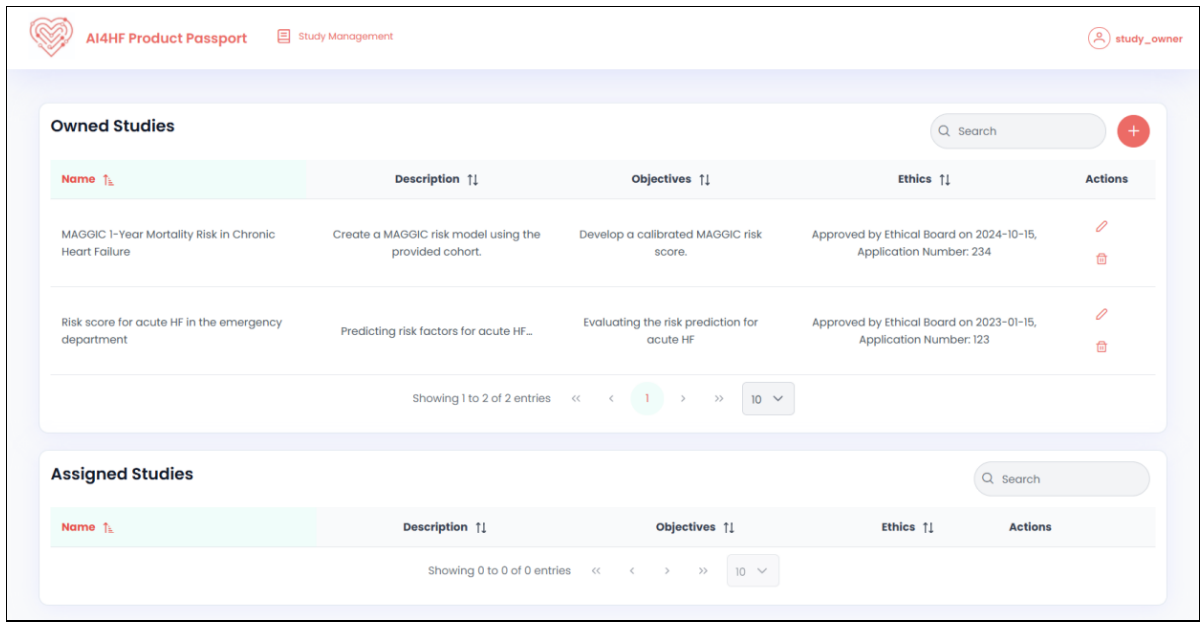

*Figure 2: Study Management Page*

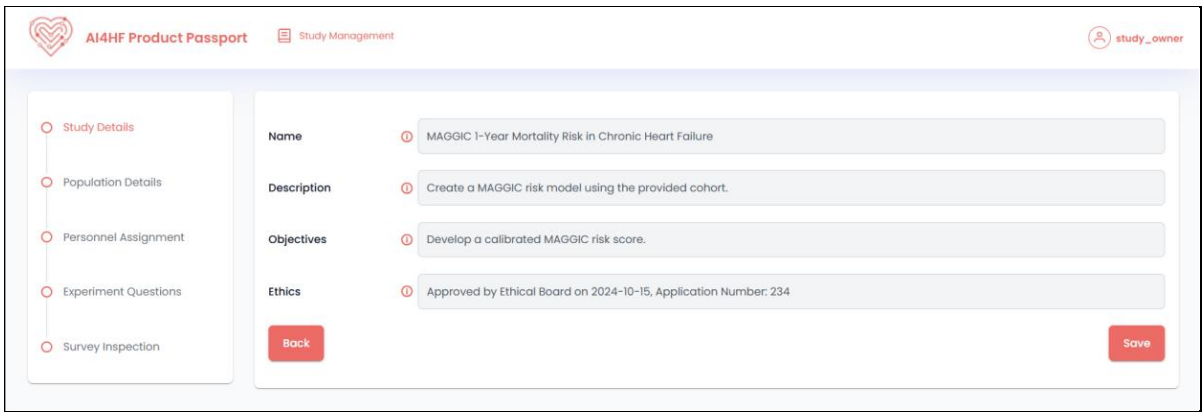

*Figure 3: Study Creation Page*

*Figure 4: Study Selection Page*

## 4. Feature Management

The Feature Management interface supports the work of the Data Engineer allows the Data Engineer to define, organize, and manage individual features and feature sets used in machine learning models within a study (Figure 5). Its structured layout simplifies the process of curating study-specific features that align with the requirements of the AI workflow (Figure 6).

*Figure 5: Feature Management Page*

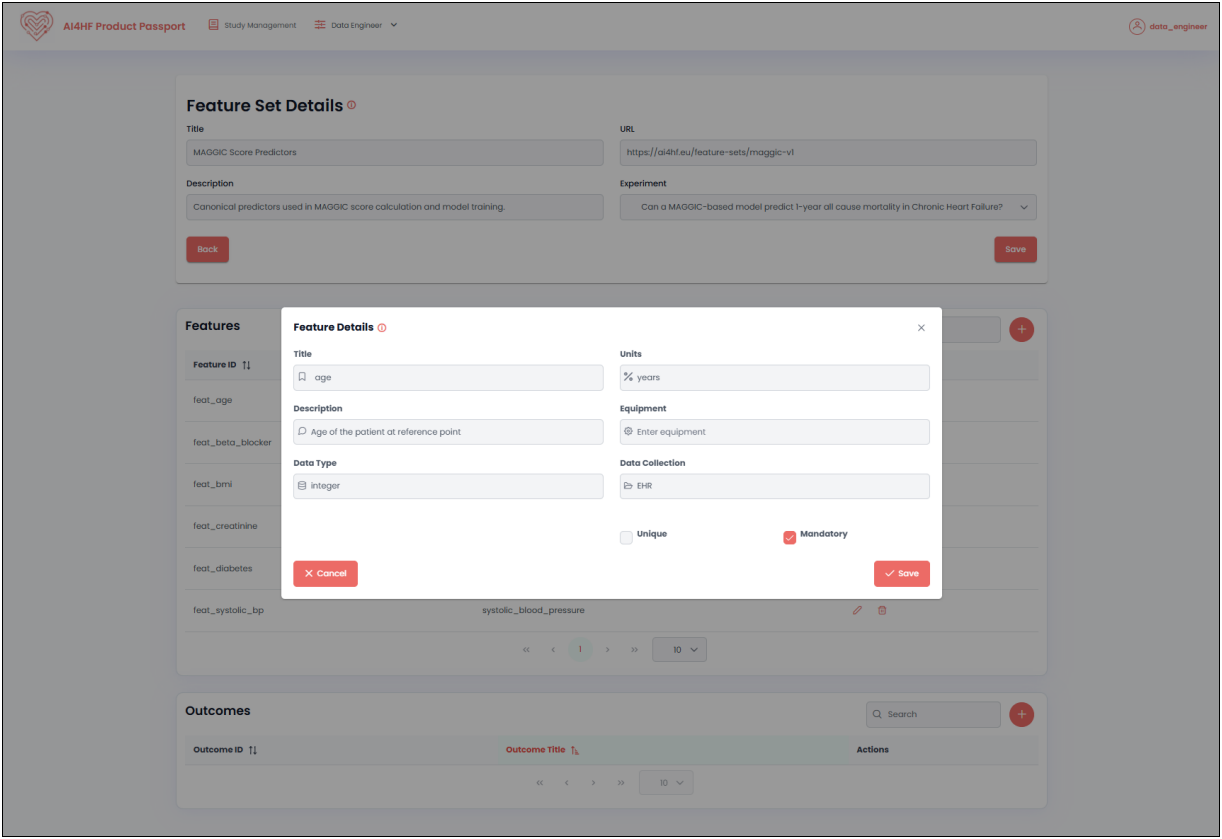

*Figure 6: Feature Creation Page*

## 5. Dataset Management

The Data Management interface supports the work of the Data Engineer by providing functionality to record detailed metadata about datasets, including their characteristics and the transformations applied to them. It allows the Data Engineer to track learning datasets and assign specific transformation steps as needed (Figure 7). This structured approach helps maintain clarity and consistency in data preparation across different studies.

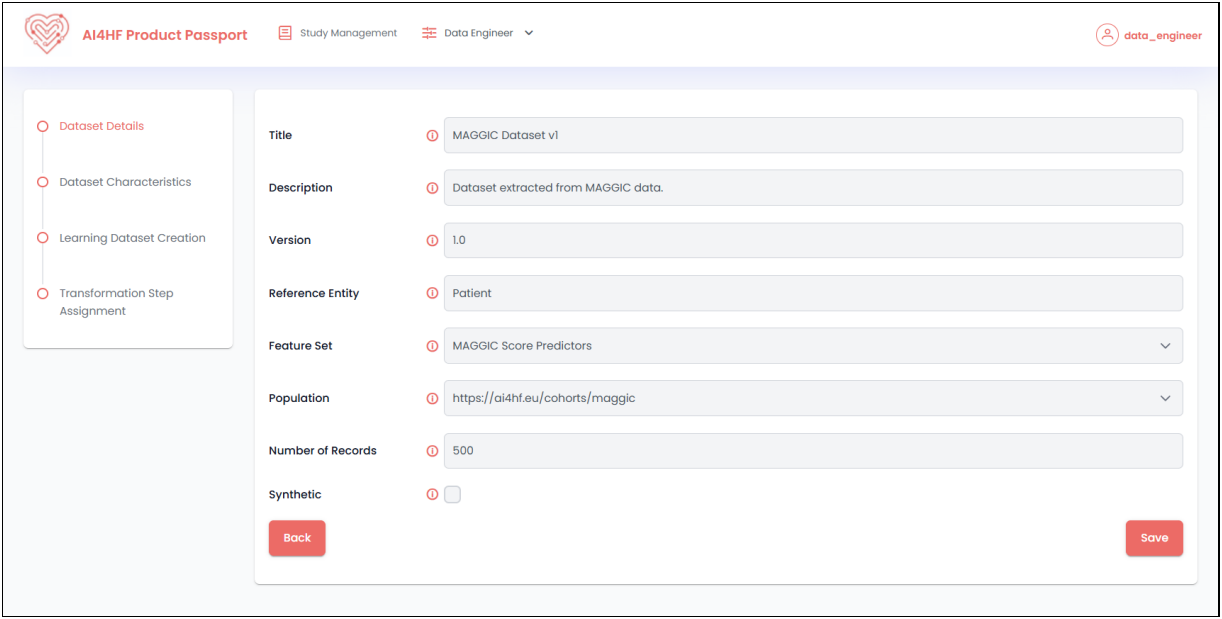

*Figure 7: Dataset Management Page*

## 6. Parameter Management

The Parameter Management interface is designed to support the Data Scientist and through this interface, the Data Scientist can define and manage key model parameters, including algorithm configurations and hyperparameters, enabling a controlled and transparent model development process (Figure 8 and Figure 9).

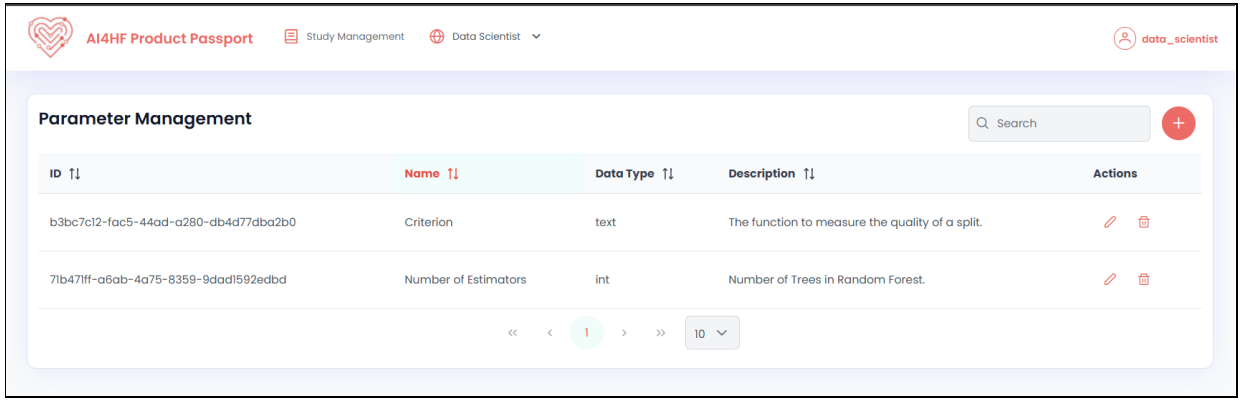

*Figure 8: Parameter Management Page*

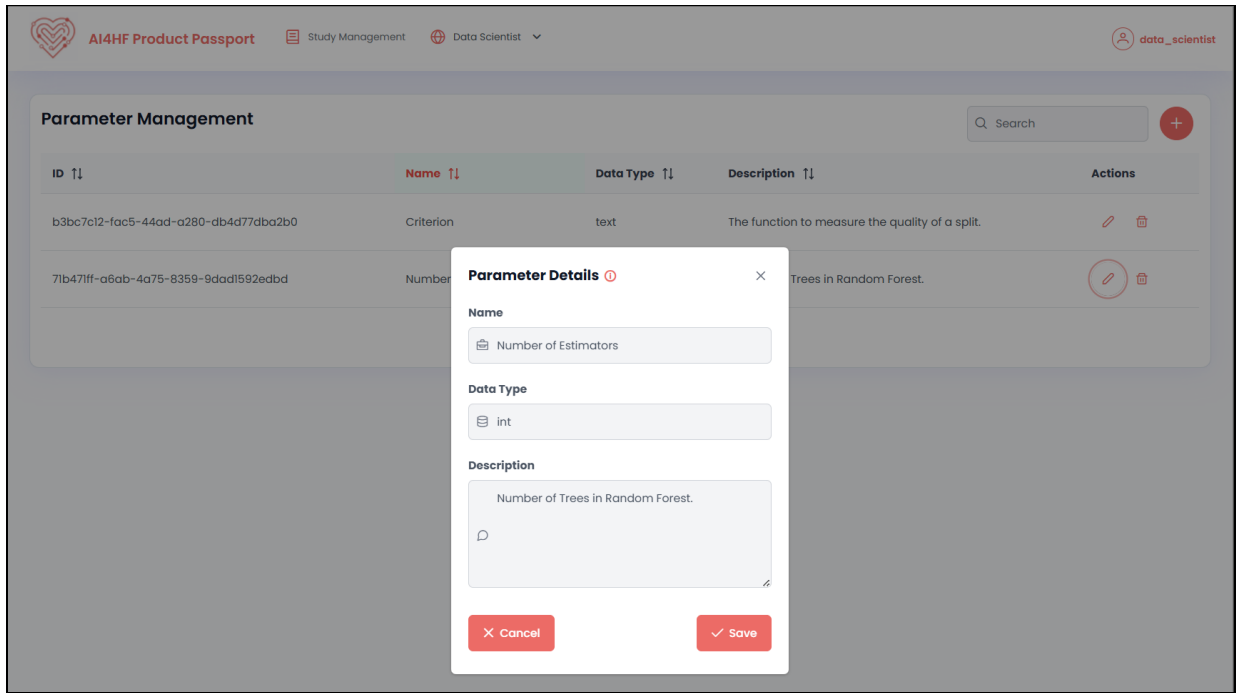

*Figure 9: Parameter Details Page*

## 7. Learning Process Management

The Learning Process Management interface supports the responsibilities of the Data Scientist. This interface allows the Data Scientist to manage process details, assign relevant datasets, define learning stages, and document process parameters (Figure 10). It provides a structured way to track the progression of ML workflows, helping to maintain consistency and clarity throughout the model development lifecycle (Figure 11).

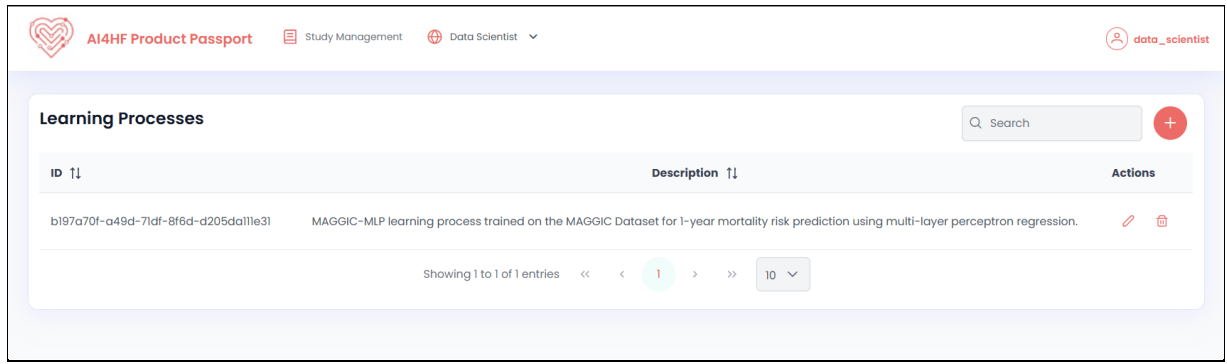

*Figure 10: Learning Process Management Page*

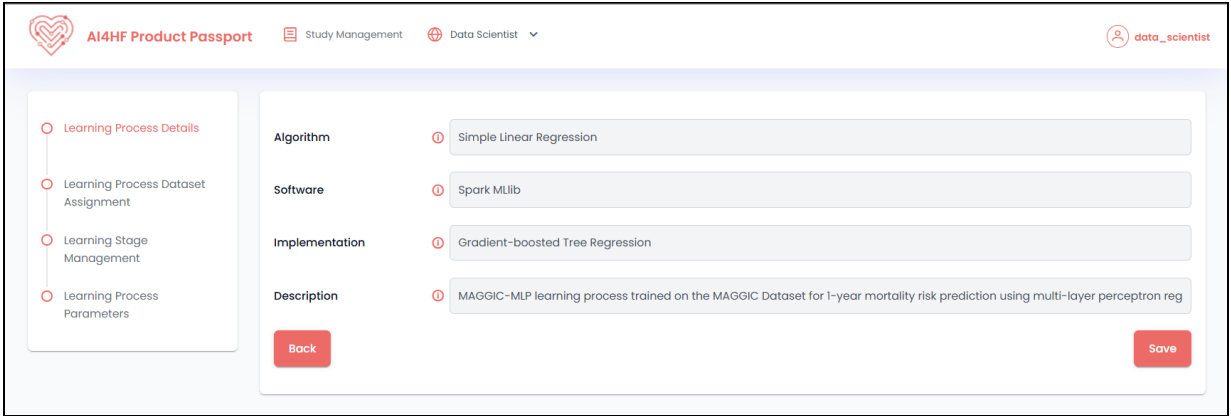

*Figure 11: Learning Process Details Page*

## 8. Model Management

The Model Management interface assists the Data Scientist in managing AI models by recording all relevant details about each machine learning model developed. It also links each model to its corresponding learning process, creating a clear connection between the training workflow and the resulting model (Figure 12 and Figure 13). This helps maintain traceability during model evaluation and deployment.

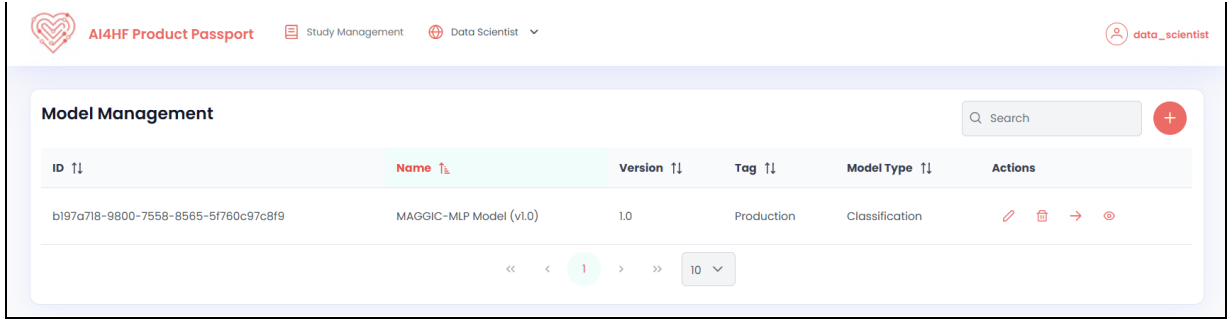

*Figure 12: Model Management Page*

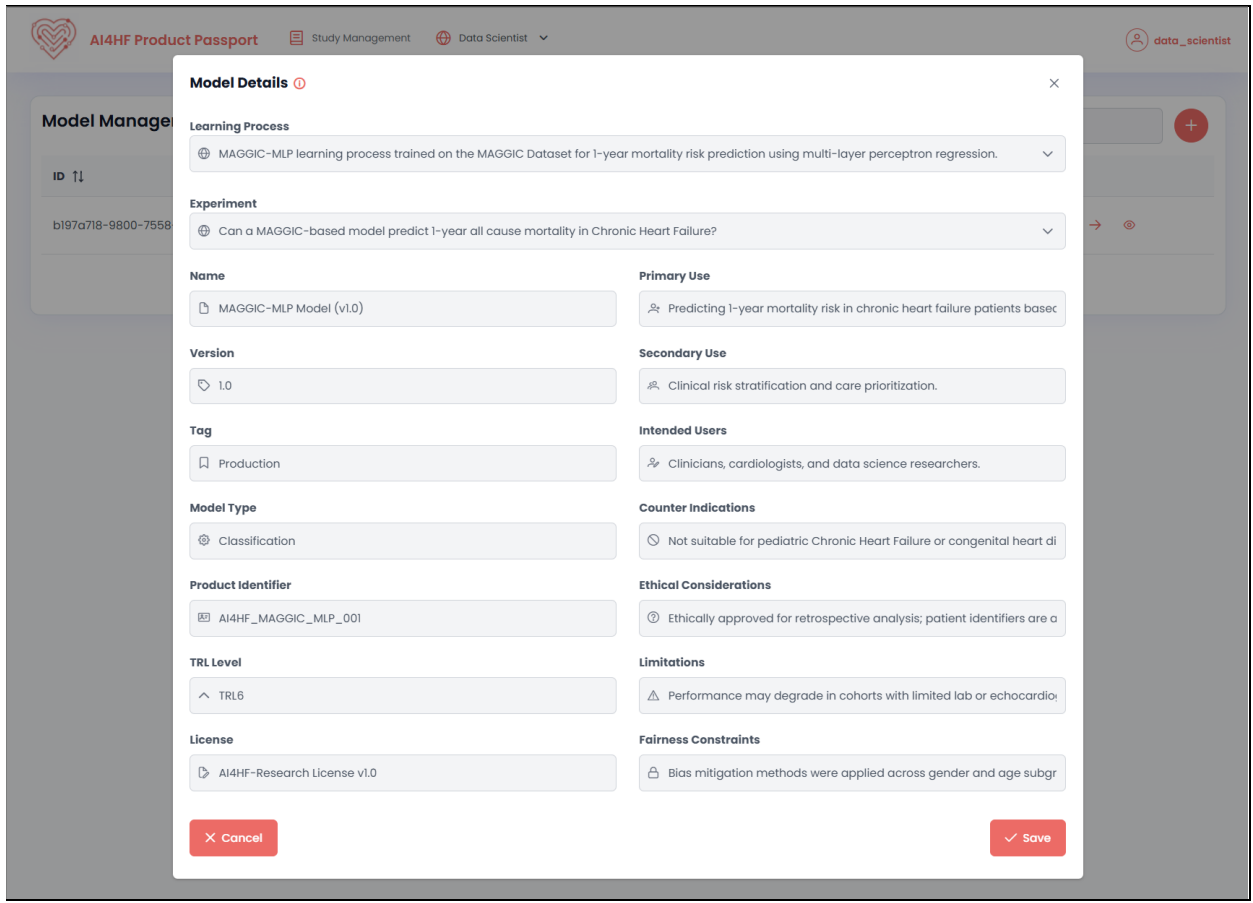

*Figure 13: Model Details Page*

## 9. Deployment Management

The Deployment Management interface supports the work of the ML Engineer. This interface tracks the deployment of ML models by recording environment details, deployment status, and any identified issues (Figure 14 and Figure 15). It provides a centralized view to monitor and manage the operational phase of AI solutions effectively.

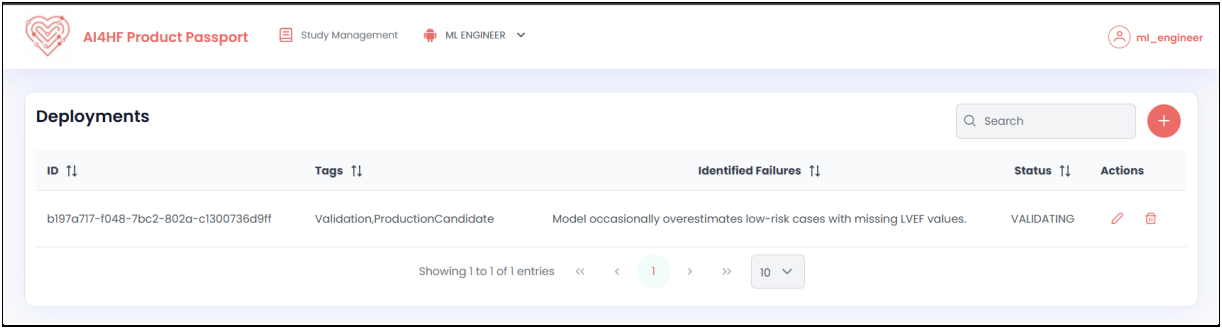

*Figure 14: Deployment Management Page*

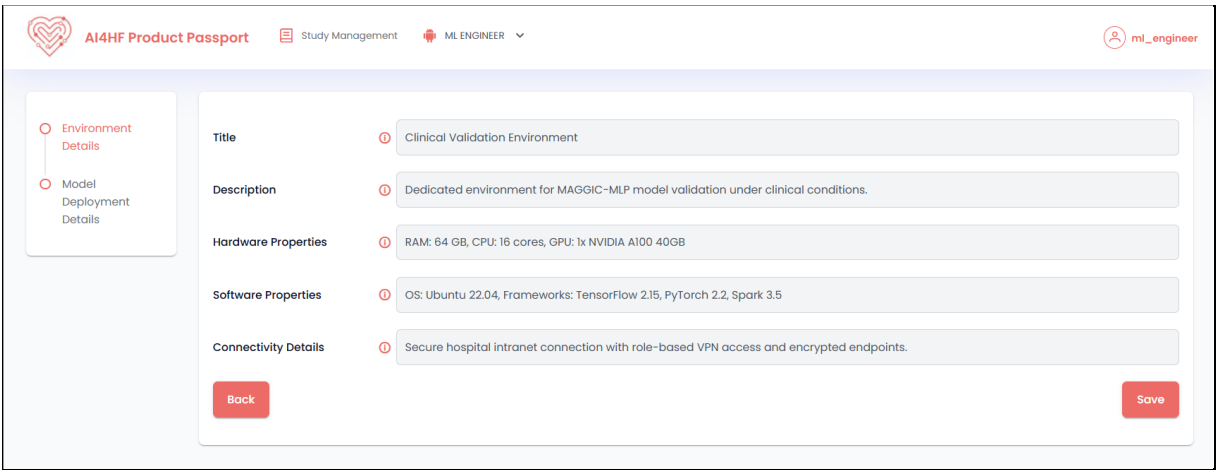

*Figure 15: Deployment Environment Page*

## 10. Passport Management

The Passport Management interface is designed to support the Quality Assurance Specialist, who is responsible for managing and generating AI Product Passports. This interface offers an overview of the AI model lifecycle, presenting detailed information about model development, deployment, and evaluation. By documenting these aspects in a structured and accessible format, the page promotes transparency and traceability, aligning with regulatory and compliance requirements (Figure 16, Figure 17, Figure 18, Figure 19, Figure 20 and **Error! Reference source not found.**).

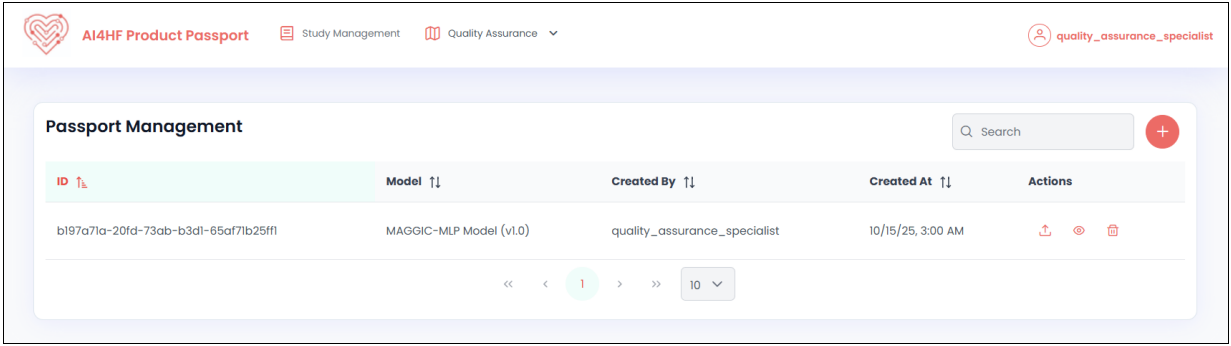

*Figure 16: Passport Management Page*

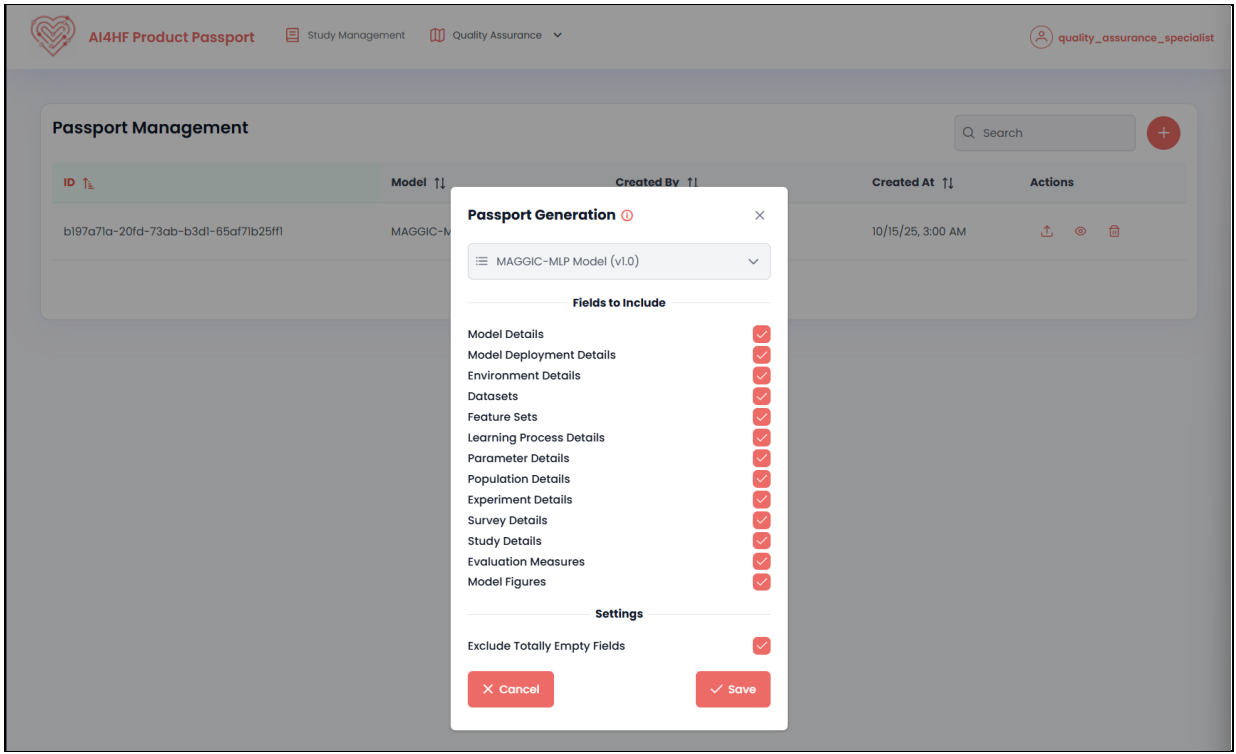

*Figure 17: Passport Level of Detail Selection Functionality*

# AI4HF Product Passport

## Model Details

| | |
|---|---|
| Model Name | MAGGIC-MLP Model (v1.0) |
| Model Version | 1.0 |
| Model Type | Classification |
| Product Identifier | AI4HF_MAGGIC_MLP_001 |
| Owner | Amsterdam UMC |
| TRL Level | TRL6 |
| License | AI4HF-Research License v1.0 |
| Primary Use | Predicting 1-year mortality risk in chronic heart failure patients based on MAGGIC feature set. |
| Secondary Use | Clinical risk stratification and care prioritization. |
| Intended Users | Clinicians, cardiologists, and data science researchers. |
| Counter Indications | Not suitable for pediatric Chronic Heart Failure or congenital heart disease populations. |
| Ethical Considerations | Ethically approved for retrospective analysis; patient identifiers are anonymized. |
| Limitations | Performance may degrade in cohorts with limited lab or echocardiographic data. |
| Fairness Constraints | Bias mitigation methods were applied across gender and age subgroups. |

## Model Deployment Details

| | |
|---|---|
| Deployment Tags | Validation,ProductionCandidate |
| Deployment Status | VALIDATING |
| Identified Failures | Model occasionally overestimates low-risk cases with missing LVEF values. |

## Environment Details

| | |
|---|---|
| Environment Title | Clinical Validation Environment |
| Environment Description | Dedicated environment for MAGGIC-MLP model validation under clinical conditions. |
| Hardware Properties | RAM: 64 GB, CPU: 16 cores, GPU: 1x NVIDIA A100 40GB |
| Software Properties | OS: Ubuntu 22.04, Frameworks: TensorFlow 2.15, PyTorch 2.2, Spark 3.5 |
| Connectivity Details | Secure hospital intranet connection with role-based VPN access and encrypted endpoints. |

## Feature Sets

| FeatureSet Title | Feature Set Description | FeatureSet URL | Created At | Created By | Last Updated At | Last Updated By |
|---|---|---|---|---|---|---|
| MAGGIC Score Predictors | Canonical predictors used in MAGGIC score calculation and model training. | https://ai4hf.eu/feature-sets/maggic-v1 | Jan 21, 1970, 11:01:26 AM | data_engineer | Jan 21, 1970, 11:01:26 AM | data_engineer |

| Feature Title | Feature Description | Feature Data Type | Feature Mandatory | Feature Units | Feature Equipment | Feature Data Collection |
|---|---|---|---|---|---|---|
| gender | Gender of the patient | string | Yes | N/A | N/A | EHR |
| age | Age of the patient at reference point | integer | Yes | years | N/A | EHR |
| nyha | The latest value of the New York Heart Assessment as LOINC Code | string | Yes | N/A | N/A | EHR/Clinical assessment |
| systolic_blood_pressure | Average systolic blood pressure (mmHg) over the 3 years preceding reference time point | decimal | Yes | mmHg | N/A | Vitals |
| bmi | Average Body Mass Index (kg/m²) over the 3 years preceding reference time point | decimal | Yes | kg/m2 | N/A | Vitals |
| | Creatinine [Mass/volume] in | | | | | |

*Figure 18: Example AI Product Passport*

| | | | | | | |
|---|---|---|---|---|---|---|
| creatinine | Creatinine [mass/volume] in Serum or Plasma (mg/L) – 3-year average | decimal | Yes | mg/L | N/A | Lab results |
| heart_failure_ge_18_months | Heart failure diagnosed ≥ 18 months before reference point | boolean | No | N/A | N/A | EHR/conditions |
| chronic_obstructive_pulmonary_disease | Presence of chronic obstructive pulmonary disease (COPD) | boolean | No | N/A | N/A | EHR/conditions |
| diabetes | Presence of diabetes mellitus | boolean | No | N/A | N/A | EHR/conditions |
| beta_blocker_use | Administration of beta-blocker medication | boolean | No | N/A | N/A | Medication administration |
| ace_inhibitor_or_arb_use | Administration of ACE inhibitor or ARB medication | boolean | No | N/A | N/A | Medication administration |
| lvef | Most recent left ventricular ejection fraction before reference point | decimal | Yes | percent | N/A | Echocardiography |
| smoker | Most recent recorded smoking status before reference point (1 = current smoker, 0 = former/never) | boolean | No | N/A | N/A | EHR/Questionnaire |

## Datasets

| Dataset Title | Dataset Description | Dataset Version | Dataset Reference Entity | Dataset Number of Records | Dataset Synthetic |
|---|---|---|---|---|---|
| MAGGIC Dataset v1 | Dataset extracted from MAGGIC data of ABC Hospital (Turkey). | 1.0 | Patient | 500 | No |

**Learning Dataset Description**

Finalized learning dataset derived from MAGGIC Dataset v1 for 1-year mortality prediction after planned transformations.

## Learning Processes

**Learning Process Description**

MAGGIC-MLP learning process trained on the MAGGIC Dataset for 1-year mortality risk prediction using multi-layer perceptron regression.

| Learning Stage Name | Learning Stage Description | Dataset Percentage |
|---|---|---|
| Training | Model training stage using 70% of the MAGGIC Dataset for multi-layer perceptron calibration. | 70% |
| Validation | Model validation stage using 15% of the MAGGIC Dataset to tune hyperparameters and prevent overfitting. | 15% |
| Testing | Final evaluation on the remaining 15% of the MAGGIC Dataset for performance assessment. | 15% |

## Population Details

| Population URL | Population Description | Population Characteristics |
|---|---|---|
| https://ai4hf.eu/cohorts/maggic | Patients with heart failure meeting MAGGIC inclusion criteria. | The study population comprised 500 participants, evenly distributed between males and females, with mean age being 28. |

## Experiment Details

**Research Question**

Can a MAGGIC-based model predict 1-year all cause mortality in Chronic Heart Failure?

## Study Details

*Figure 19: Example AI Product Passport (cont.)*

| | |
|---|---|
| **Study Name** | MAGGIC 1-Year Mortality Risk in Chronic Heart Failure |
| **Study Description** | Create a MAGGIC risk model using the provided cohort. |
| **Study Ethics** | Approved by Ethical Board on 2024-10-15, Application Number: 234 |
| **Study Objectives** | Develop a calibrated MAGGIC risk score. |

## Evaluation Measures

| Name | Data Type | Value | Description |
|---|---|---|---|
| AUC | float | 0.89 | Area under the ROC curve for 1-year mortality prediction using MAGGIC-MLP. |
| Accuracy | float | 0.83 | Overall classification accuracy of the MAGGIC-MLP model. |
| F1-Score | float | 0.81 | F1-score reflecting precision and recall balance for the MAGGIC-MLP model. |

## Model Figures

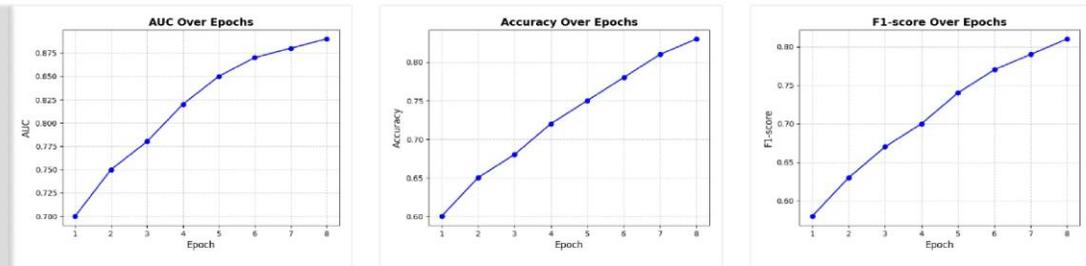

*Figure 20: Example AI Product Passport (cont.)*

# 11. Python Software Library

The Software API for automating data collection is designed as a Python library to simplify the process of gathering and submitting information about ML models directly into the AI Product Passport during the model generation and evaluation phase. This API enables developers to seamlessly integrate data collection into their existing workflows to reduce the manual effort involved in documenting AI models by automating the metadata collection process. As developers work on their models, this library provides functions that they can call to send updated information directly to the AI Product Passport. This ensures that the passport always contains up-to-date data about each model, without requiring developers to interrupt their work to manually update records (Figure 21 and Figure 22).

```python
import requests
import json
import jwt
from typing import Dict, Any
from ai4hf_passport_models import *

class BaseMetadataCollectionAPI:
    """
    Base class for interacting AI4HF Passport Server.
    """

    def __init__(self, passport_server_url: str, study_id: str, experiment_id: str, organization_id: str, username: str,
                 password: str):
        """
        Initialize the API client with authentication and study details.
        """
        self.passport_server_url = passport_server_url
        self.study_id = study_id
        self.experiment_id = experiment_id
        self.organization_id = organization_id
        self.username = username
        self.password = password
        self.token = self._authenticate()

    def __str__(self):
        return json.dumps({"passport_server_url": self.passport_server_url, "study_id": self.study_id,
                           "experiment_id": self.experiment_id,
                           "organization_id": self.organization_id, "username": self.username,
                           "password": self.password, "token": self.token})

    def _refreshTokenAndRetry(self, response, headers, payload, url):
        """
        If token is expired, refresh token and retry

        :param response: Response object from previous request.
        :param headers: Headers object from previous request.
        :param payload: Payload object from previous request.
        :param url: The url to sent.

        :return response: Response algorithm object from the server.
        """

        if response.status_code == 401:  # Token expired, refresh and retry
            self.token = self._authenticate()
            headers["Authorization"] = f"Bearer {self.token}"
            return requests.post(url, json=payload, headers=headers)
        else:
            return response
```

*Figure 21: Code example from Python Software Library*

```python
# ******** * * *   *   *    *     *    *    *    *  * * * * * * **************
# Construct an api client for interacting with AI4HF passport server

api_client = TorchMetadataCollectionAPI(
        passport_server_url="http://localhost:80/ai4hf/passport/api",
        study_id="initial_study",
        organization_id="initial_organization",
        username="data_scientist",
        password="data_scientist"
    )

# Provide learning stages
learning_stages = [
    LearningStage(learningStageType = LearningStageType.TRAINING,
                  datasetPercentage = int(100*params.train_size)),
    LearningStage(learningStageType = LearningStageType.TEST,
                  datasetPercentage = int(100*params.test_size)),
    LearningStage(learningStageType = LearningStageType.VALIDATION,
                  datasetPercentage = int(100*params.val_size))
    ]

with open("metrics.json", "r") as f:
    metrics_file = json.load(f)
# {'tempauroc': 0.5018877581090273, 'tempprc': 0.5009819110168369, 'loss': 2.470166275946147}

# Provide evaluation measures
evaluation_measures = [
    EvaluationMeasure(EvaluationMeasureType.MAE,
                      value = str(metrics_file["loss"])),
    EvaluationMeasure(EvaluationMeasureType.ROC,
                      value = str(metrics_file["tempprc"])),
    EvaluationMeasure(EvaluationMeasureType.AUC,
                      value = str(metrics_file["tempauroc"]))
    ]
"""
    loss defined :
    loss_fct = SurvODELoss(reduction='mean')
"""
# Provide model details
model_info = Model(
    name = "test",
    modelType = "MLP",
    version = "1.0")

# Call this function with your model object
api_client.submit_results_to_ai4hf_passport(model, learning_stages, evaluation_measures, model_info)

## COSAS POR AHCER:
# Hacer variable el directorio del lib passport
# personalizar el lerning stage y el evaluation measures
```

*Figure 22: An example code portion that directly sends model metadata to AI Product Passport server directly from development environment by using Python Software Library*